\documentclass[11pt]{article}
\usepackage{latexsym,cite,amssymb}
\usepackage{times}
\makeatletter

\@addtoreset{equation}{section} \makeatother

\input amssym.def
\input amssym.tex

\newcommand{\half}{{{\textstyle\frac{1}{2}}}}

\newcommand{\be}{\begin{equation}}
\newcommand{\ee}{\end{equation} }
\newcommand{\beqa}{\begin{eqnarray} }
\newcommand{\eeqa}{\end{eqnarray} }
\newcommand{\ba}{\begin{array}}
\newcommand{\ea}{\end{array}}

\newcommand{\so}{\mathbf{so}}
\newcommand{\SO}{\mathbf{SO}}

\newcommand\Tr{{\rm Tr}}

\newcommand\rd{{\rm d}}

\newcommand\cL{{\cal L}}
\newcommand\cM{{\cal M}}

\newcommand\cV{{\cal V}}

\newcommand\hd{p}

\newcommand\NaGo{{\rm \scriptscriptstyle{N.G.}}}

\newcommand\stSUSY{{\rm \scriptscriptstyle{spacetime}}}
\newcommand\WZ{{\rm \scriptscriptstyle{Wess-Zumino}}}
\newcommand\wSUSY{{\rm \scriptscriptstyle{worldline}}}
\newcommand\NB{{\rm \scriptscriptstyle{N.B.}}}

\newcommand\diff{{}}

\newcommand\dis{\displaystyle}
\newcommand\bosonic{{{{\rm{bosonic}}}}}
\newcommand\multistring{{{\rm{string}}}}

\begin{document}
\begin{titlepage}
\title{
\vskip 2cm
Partonic description of a supersymmetric  $p$-brane\\~\\}
\author{\sc
Kanghoon Lee${}^{\sharp}$ \mbox{~~\,\,}and\mbox{\,\,~~} Jeong-Hyuck Park${}^{\dagger}$}
\date{}
\maketitle \vspace{-1.0cm}
\begin{center}
~~~\\
${}^{\sharp}$Department of Physics,   Yonsei University, Shinchon-dong, Seodaemun-gu, Seoul 120-749, Korea\\
\texttt{lkh@phya.yonsei.ac.kr}\\
~{}\\
${}^{\dagger}$Department of Physics, Sogang University, Shinsu-dong, Mapo-gu, Seoul 121-742, Korea\\
\texttt{park@sogang.ac.kr}
~{}\\
~~~\\
~~~\\
\end{center}
\begin{abstract}
\vskip0.5cm
\noindent
We consider  supersymmetric extensions of a recently proposed    partonic description of  a bosonic $p$-brane  which reformulates the Nambu-Goto action as an  interacting  multi-particle action with    Filippov-Lie algebra gauge  symmetry.   We  construct  a worldline  supersymmetric  action by postulating, among others,     a $p$-form fermion.   Demanding a local worldline supersymmetry rather than the full worldvolume supersymmetry, we circumvent a known  no-go theorem against  the construction of a   Ramond-Neveu-Schwarz  supersymmetric action for a $p$-brane of ${p>1}$.   We also derive a    spacetime supersymmetric  Green-Schwarz  extension    from   the preexisting    kappa-symmetric action.  
\end{abstract}

{\small
\begin{flushleft}
~~~~~~~~\textit{PACS}: 11.25.-w, 11.30.Pb \\
~~~~~~~~\textit{Keywords}: parton,  $p$-brane, supersymmetry 
\end{flushleft}}
\thispagestyle{empty}
\end{titlepage}
\newpage

\tableofcontents 
\section{Introduction}

Supersymmetry in string theory is two fold:  one on the string  worldsheet   by the  Ramond-Neveu-Schwarz  formalism and the other one  on the  spacetime  by the Green-Schwarz  formalism.
It is known that these two approaches are equivalent, at least for ten-dimensional Minkowskian spacetime. One may attempt to extend the two formalisms  to a $p$-brane with $p>1$  \textit{i.e.~}an extended object over   $p$-spatial dimensions. Indeed, the extension of the Green-Schwarz covariant superstring action to
a $p$-brane  is possible for $p\leq 5$~\cite{Achucarro:1987nc}.  The resulting action is invariant  under not only   the spacetime supersymmetry but also a fermionic gauge symmetry called kappa-symmetry, such that the on-shell  Bose and Fermi degrees of freedom are equal. 
On the other hand,  for a $p$-brane with $p>1$,  the   Ramond-Neveu-Schwarz  extension to  the corresponding   Nambu-Goto action reformulated by    an auxiliary worldvolume metric~\cite{Deser:1976rb,Brink:1976sc,Howe:1977hp}\footnote{This action is often dubbed ``Polyakov" action.}   is known  impossible: in Ref.\cite{Bergshoeff:1988ui} it was shown that  the worldvolume supersymmetric extension  requires the existence of the Einstein-Hilbert term for the worldvolume metric such that the metric is no longer auxiliary and the connection to the Nambu-Goto action is lost. \\

\noindent Recently,  a partonic description of a bosonic  $p$-brane was proposed in Ref.\cite{Park:2008qe}.  With an embedding of  ${(p{+1})}$-dimensional worldvolume coordinates  into  $D$-dimensional target spacetime, $X^{M}(\tau,\sigma^{i})$ where and henceforth  $i=1,\cdots, p$ and $M=0,1,\cdots,D{-1}$,  the proposed action assumes the form:
\be
\dis{S_{\bosonic}=
\int\!\rd\tau~\Tr\Big(\half D_{\tau}X^{M}D_{\tau}X_{M}
-\textstyle{\frac{1}{\,2\hd!}}\{X^{M_{1}},X^{M_{2}},\cdots,X^{M_{\hd}}\}_{\NB}
\{X_{M_{1}},X_{M_{2}},\cdots,X_{M_{\hd}}\}_{\NB}\Big)\,.}
\label{actionbosonic}
\ee
The action  contains two kinds of auxiliary fields:  the inverse of an einbein  $\varphi$ and  a gauge connection $A^{i}_{\tau}$. The former defines the trace inside the  action,
\be
\dis{\Tr\big(~\cdot~\big):=\int \rd^{p}\sigma}\,\big(\varphi~\cdot~\big)\,,
\label{Trace}
\ee
and  the   Nambu bracket\footnote{As usual,  $\epsilon^{i_{1}i_{2}\cdots i_{{\hd}}}$ is   the totally anti-symmetric $p$-dimensional  tensor density  with the normalization  $\epsilon^{12\cdots p}=1$.}~\cite{Nambu:1973qe},
\be
\{X^{M_{1}},X^{M_{2}},\cdots,X^{M_{\hd}}\}_{\NB}
:=\varphi^{-1}\epsilon^{i_{1}i_{2}\cdots i_{{p}}}\partial_{i_{1}}X^{M_{1}}\partial_{i_{2}}X^{M_{2}}\cdots\partial_{i_{\hd}}X^{M_{p}}\,,
\ee
while the latter sets  the covariant derivative to be
\be
D_{\tau}X^{M}:=\partial_{\tau}X^{M}-A_{\tau}^{~i}\partial_{i}X^{M}\,.
\ee
In fact,    $\varphi$ and $A^{i}_{\tau}$ may be identified with the ``lapse" and ``shift" Lagrange multipliers of the Nambu-Goto  action in the canonical formalism~\cite{Henneaux:1983um}.  Characteristic features of the above action are  \cite{Park:2008qe}:
\begin{itemize}
\item  Integrating out the auxiliary fields, \textit{i.e.~}replacing them by their on-shell values, reduces the action to the standard Nambu-Goto action, as in \cite{Deser:1976rb,Brink:1976sc,Howe:1977hp} or \cite{Schild:1976vq}.
\item  The action is manifestly  spacetime Lorentz invariant, despite the  similarity  to the  
 light-cone gauge fixed  actions in \cite{Hoppe,Bergshoeff:1988hw}.
 \item Though not manifest, the action enjoys the full $(p{+1})$-dimensional worldvolume diffeomorphism. 
 \item The number of  auxiliary component fields is ${p+1}$, and hence the  worldvolume diffeomorphism can fix  them completely, such as $\varphi\equiv1$ and $A^{i}_{\tau}\equiv0$.\footnote{\textit{c.f.~}``Polyakov" action where the   
number of  auxiliary component fields is ${\half(p+1)(p+2)}$ such that for $p>1$ they can not be  gauge fixed completely. }
 
 \item Partial gauge fixing as  $\varphi\equiv1$ and $\partial_{i}A^{i}_{\tau}\equiv0$ breaks the worldvolume diffeomorphism  to a  volume preserving $p$-dimensional diffeomorphism. For a compact $p$-brane this  leads to  a quantum mechanical system based on Filippov-Lie $p$-algebra.\footnote{Relates works are the Bagger-Lambert-Gustavsson description of  multiple M2-branes \textit{via}  Filippov-Lie three-algebra~\cite{BL,Gustavsson:2007vu}.}

 \item A physical picture behind  the reformulation is to describe a single (compact) brane as a collection of interacting 
 multi-particles, and hence the title of this paper: \textit{partonic description of a $p$-brane}.\footnote{Related works include BFSS $\cM$-theory  matrix model~\cite{Banks:1996vh} and Myers' effect~\cite{Myers:1999ps} \textit{etc}.  See also ~\cite{Bergshoeff:1988hw,Yang:1998qd}.}
  
 \item One may consider  implementing  a \textit{worldline} supersymmetry, rather than the full $(p{+1})$-dimensional worldvolume supersymmetry.
 \end{itemize}

In the present paper, we focus on exploring the last property.  In the multi-particle description of a (compact)  $p$-brane,      the temporal worldline direction is singled out from the full $(p{+1})$-dimensional worldvolume.    Demanding  a local  supersymmetry along the worldline we may circumvent the aforementioned no-go theorem against  the construction of a    Ramond-Neveu-Schwarz  supersymmetric action for a $p$-brane with ${p>1}$.\footnote{For earlier proposals  to circumvent the no-go theorem, we refer \cite{Lindstrom:1988az,Castro:2002ra}. } \\

Filippov-Lie $p$-algebra appears as a natural generalization of Lie-algebra \textit{i.e.~}two-algebra. While Lie algebra has been extensively studied ever since the inception  of  Yang-Mills theory,   Filippov-Lie $p$-algebra with $p>2$ had  not been much explored until 2007 when Bagger-Lambert and Gustavsson employed  Filippov-Lie three-algebra with the aim to    describe    multiple M2-branes~\cite{BL,Gustavsson:2007vu}. In the present paper we shall construct supersymmetric gauge  models based on arbitrary Filippov-Lie $p$-algebras. \\

\noindent The organization of the rest of the paper is as follows.  In section~\ref{secReview}, we  review the bosonic action (\ref{actionbosonic}) with some details, including the full $(p{+1})$-dimensional worldvolume diffeomorphism and the Filippov-Lie $p$-algebra regularization.   The worldline  supersymmetric  action is constructed in section~\ref{secWorldline}.  We first present a  foliation preserving,  diffeomorphism invariant and  locally supersymmetric Ramond-Neveu-Schwarz  action. After gauge fixing we also obtain an action with a global supersymmetry.  A crucial ingredient in our worldline supersymmetric extension  is to postulate a $p$-form fermion,  in addition to a one-form fermion and a gravitino.  In section~\ref{secSpacetime},   we derive a  spacetime supersymmetric  Green-Schwarz extension    from     
the known  kappa-symmetric action.  We write down the proper transformation of the auxiliary fields which will ensure all the symmetries of the spacetime supersymmetric Nambu-Goto action to persist in our reformulation.   Section~\ref{secDiscussion} contains our discussion and Appendix carries some useful identities.\\


\section{More on the bosonic action (\ref{actionbosonic})\label{secReview}}
In this section we review, from Ref.\cite{Park:2008qe},  some properties of the bosonic action (\ref{actionbosonic}) which are relevant to our main results of the supersymmetrization.  

With a ${p\times p}$ matrix defined by
\be
V_{ij} := \partial_{i}X^{M}\partial_{j} X_{M}\,,
\ee
utilizing an identity,
\be
\varphi^{-2}\det V\!=\textstyle{\frac{1}{\hd!}}\{X^{M_{1}},X^{M_{2}},\cdots,X^{M_{\hd}}\}_{\NB}
\{X_{M_{1}},X_{M_{2}},\cdots,X_{M_{\hd}}\}_{\NB}\,,
\label{potential}
\ee
the bosonic action (\ref{actionbosonic}) can be rewritten as
\be
\dis{S_{\bosonic}=
\int\rd\tau\rd^{p}\sigma~\Big(\half \varphi D_{\tau}X^{M}D_{\tau}X_{M}
-\half\varphi^{-1}\det V\Big)\,.}
\label{startingaction}
\ee
The on-shell values of  the auxiliary fields are
\be
\ba{ll}
A_{\tau}^{~i}\equiv \partial_{\tau}X^{M}\partial_{j} X_{M}V^{-1ji}\,,~&~
\varphi\equiv\sqrt{-\det V/(D_{\tau}X^{M}D_{\tau}X_{M})}\,.
\ea
\label{onshellaux}
\ee
Substituting  these into the action~(\ref{startingaction}),  one recovers the Nambu-Goto action~\cite{NambuGoto},
\be
\ba{lll}
S_{\bosonic}&~~\Longrightarrow~~& S_{\NaGo}=\displaystyle{-\int}
\rd\tau\rd^{p}\sigma~\sqrt{-\det\left(\partial_{\mu}X^{M}\partial_{\nu}X_{M}\right)}\,,
\ea
\label{NGp}
\ee
where and henceforth $\mu,\nu$ are the full  ${(p+1)}$-dimensional worldvolume coordinate indices running from zero to $p$.   The   worldvolume   diffeomorphism is realized in  rather nontrivial fashion:
\be
\ba{l}
\delta_{\diff}X^{M}=\upsilon^{\mu}\partial_{\mu}X^{M}\,,\\
\delta_{\diff}\varphi=\partial_{\mu}\!\left(\varphi\upsilon^{\mu}\right)
-2\varphi D_{\tau}\upsilon^{\tau}\,,\\
\delta_{\diff} A_{\tau}^{~i}=D_{\tau}\upsilon^{i}
-\varphi^{-2}\partial_{j}\upsilon^{\tau}V^{-1 ji}\det V +
\left(D_{\tau}\upsilon^{\tau}+\upsilon^{\mu}\partial_{\mu}\right) A_{\tau}^{~i}\,,
\ea
\label{diffgend1}
\ee
where $\upsilon^{\mu}$ is a local parameter having  an arbitrary dependence on $\tau$ and $\sigma^{i}$.    
In general, a symmetry of a given  action persists after any reformulation by   auxiliary fields: we can always assign  transformations to the auxiliary fields such that the symmetry is preserved~\cite{Bergshoeff:1988ui}. The above transformation (\ref{diffgend1}) is an explicit example of this general statement.  \\

From
\be
\varphi\!\left(D_{\tau}Y Z+YD_{\tau}Z\right)=YZ\!
\left[\partial_{i}\!\left(\varphi A^{i}\right)-\partial_{\tau}\varphi\right]+
\partial_{\tau}\!\left(\varphi YZ\right)-\partial_{i}\!\left(\varphi A_{\tau}^{i}YZ\right)\,,
\ee
the vanishing of the following quantity,
\be
\partial_{i}\!\left(\varphi A_{\tau}^{i}\right)-\partial_{\tau}\varphi=\varphi
\!\left(\partial_{i}A_{\tau}^{i}-D_{\tau}\ln\varphi\right)\equiv 0\,,
\label{gf1}
\ee
is the sufficient and necessary condition of  the  integration by part for the covariant derivative:
\be
\dis{\int\!\rd\tau\,\Tr\!\left(\,D_{\tau}YZ\,\right)=-\int\!\rd\tau\,\Tr\!\left(\, YD_{\tau}Z\,\right)\,,}
\ee
with arbitrary $Y$ and $Z$. Under the  transformation (\ref{diffgend1}),  
\be
\ba{ll}
\delta_{\diff}\!\left(\partial_{i}A_{\tau}^{i}-D_{\tau}\ln\varphi\right)=&\!D_{\tau}^{2}\upsilon^{\tau}-
\textstyle{\frac{1}{(\hd-1)!}}\{X^{M_{1}},\cdots,X^{M_{\hd-1}},
\{X_{M_{1}},\cdots,X_{M_{\hd-1}},\upsilon^{\tau}\}_{\NB}\}_{\NB}\\
{}&+\left(D_{\tau}\upsilon^{\tau}+\upsilon^{\mu}\partial_{\mu}\right)\!\left(\partial_{i}A_{\tau}^{i}-D_{\tau}\ln\varphi\right)\,.
\ea
\label{gfdiff}
\ee
Thus, fixing the gauge (\ref{gf1}) generically  breaks 
the  worldline reparametrization to the global transformation,  $\upsilon^{\tau}=\alpha\tau+\beta$ with constant parameters $\alpha,\beta$, and reduces the   ${(p{+1})}$-dimensional worldvolume diffeomorphism to the  ${p}$-dimensional  diffeomorphism on the `space' part of the worldvolume. \\

On the other hand,  fixing the gauge $\varphi\equiv 1$ and $\partial_{i}A_{\tau}^{i}\equiv 0$, reduces the worldvolume diffeomorphism down to the $p$-dimensional volume preserving diffeomorphism that is subject to the divergence free condition, $\partial_{i}\upsilon^{i}=0$. Consequently, the volume preserving  gauge symmetry generator    as well as the covariant derivative can be represented by  the Nambu $p$-bracket: with a functional basis  $T^{a}(\sigma^{i})$, $a=1,2,3,\cdots$ for the $p$-dimensional manifold which we assume to be compact,  we have
\be
\ba{l}
\upsilon^{i}\partial_{i}=\upsilon_{a_{1}a_{2}\cdots a_{\hd-1}}\{T^{a_{1}},T^{a_{2}},\cdots,T^{a_{\hd-1}},~~~\}_{\NB}\,,\\
D_{\tau}=\partial_{\tau}-A_{\tau a_{1}a_{2}\cdots a_{\hd-1}}\{T^{a_{1}},T^{a_{2}},\cdots,T^{a_{\hd-1}},~~~\}_{\NB}\,.
\label{upDNB}
\ea
\ee
Note that here $\upsilon_{a_{1}a_{2}\cdots a_{\hd-1}}$ and $A_{\tau a_{1}a_{2}\cdots a_{\hd-1}}$ depend on $\tau$ only being  independent of the $\sigma^{i}$ coordinates. \\

As is well known (see \textit{e.g.~}\cite{Takhtajan:1993vr}), Nambu $\hd$-bracket provides an explicit  realization  of  the bracket of the Filippov-Lie  $\hd$-algebra~\cite{n-Lie}, satisfying the totally anti-symmetric property:
\be
[X_{1},\cdots,X_{i},\cdots,X_{j},\cdots,X_{\hd\,}]=-[X_{1},\cdots,X_{j},\cdots,X_{i},\cdots,X_{\hd\,}]\,,
\label{antisym}
\ee
and the Leibniz rule, also known as a fundamental identity:
\be
\left[X_{1},\cdots,X_{{\hd-1}},[Y_{1},\cdots,Y_{\hd\,}]\right]=\sum_{j=1}^{\hd}~
\left[Y_{1},\cdots,[X_{1},\cdots,X_{{\hd-1}},Y_{j\,}],\cdots,Y_{\hd\,}\right]\,.
\label{Leibniz}
\ee
In the Nambu bracket realization  of  the  Filippov-Lie algebra, we may employ the structure constant through
\be
\{T^{a_{1}},T^{a_{2}},\cdots,T^{a_{\hd}}\}_{\NB}=f^{a_{1}a_{2}\cdots a_{\hd}}{}_{b}T^{b}\,.
\label{structureconstant}
\ee
The structure constant  is then totally anti-symmetric for the upper indices and  satisfies from the Leibniz rule (\ref{Leibniz}):
\be
f^{a_{1}a_{2}\cdots a_{\hd}}{}_{c}f^{b_{1}b_{2}\cdots b_{\hd}}{}_{a_{\hd}}=\sum_{j=1}^{\hd}~
f^{a_{1}a_{2}\cdots a_{{\hd{-}1}}b_{j}}{}_{e}\,f^{b_{1}\cdots b_{{j{-}1}}eb_{{j{+}1}}\cdots b_{\hd}}{}_{c}\,.
\label{Leibnizf}
\ee
Now from (\ref{upDNB}) and (\ref{structureconstant}), expanding the dynamical variables by the  functional basis,  $X^{M}(\tau,\sigma)=X^{M}_{a}(\tau)T^{a}(\sigma)$, the covariant derivative can be rewritten as
\be
\ba{ll}
D_{\tau}X^{M}=(D_{\tau}X^{M})_{a}T^{a}\,,~~~~&~~~~(D_{\tau}X^{M})_{a}=
\textstyle{\frac{\rd~}{\rd \tau}}X^{M}_{a}-X^{M}_{b}\tilde{A}_{\tau}^{b}{}_{a}\,,
\ea
\label{covDBLG1}
\ee
where we set
\be
\tilde{A}_{\tau}^{b}{}_{a}:=A_{\tau c_{1}c_{2}\cdots c_{\hd-1}}f^{c_{1}c_{2}\cdots c_{\hd-1}b}{}_{a}\,.
\label{covDBLG2}
\ee
In this way,   the bosonic   action (\ref{actionbosonic}), (\ref{startingaction}) reduces  to a genuine quantum mechanical system with gauge symmetry based on an arbitrary   Filippov-Lie $\hd$-algebra.  It is worthwhile to note that for $p\geq 3$ the only nontrivial irreducible finite dimensional  Filippov-Lie  $\hd$-algebra is, up to  signature,  $\so(p{+1})$~\cite{FigueroaO'Farrill:2002xg,Papadopoulos:2008sk,Gauntlett:2008uf,Papadopoulos:2008gh}.   With $\tilde{v}^{b}{}_{a}:=v_{c_{1}c_{2}\cdots c_{\hd-1}}f^{c_{1}c_{2}\cdots c_{\hd-1}b}{}_{a}$, from
 (\ref{diffgend1}) and  (\ref{upDNB}), the Filippov-Lie  $\hd$-algebra  gauge transformation    is  given by
\be
\ba{l}
\delta X^{M}_{a}=X^{M}_{b}\tilde{v}^{b}{}_{a}\,,\\
\delta A_{\tau a_{1}a_{2}\cdots a_{\hd-1}}=\partial_{\tau}v_{a_{1}a_{2}\cdots a_{\hd-1}}+
(-1)^{\hd}(\hd-1)A_{\tau c[a_{1}a_{2}\cdots a_{\hd-2}}\tilde{v}^{c}{}_{a_{\hd-1}]}\,,
\ea
\label{Filippovgaugetransformation}
\ee
of which the latter induces, from (\ref{Leibnizf}),
\be
\delta \tilde{A}_{\tau}^{b}{}_{a}=\partial_{\tau}\tilde{v}^{b}{}_{a}-\tilde{v}^{b}{}_{c}\tilde{A}_{\tau}^{c}{}_{a}+
\tilde{A}_{\tau}^{b}{}_{c}\tilde{v}^{c}{}_{a}\,.
\ee
In fact, the case of $p=3$  matches with  the Bagger-Lambert-Gustavsson formalism~\cite{BL,Gustavsson:2007vu}.  Other useful relations for the bosonic action are written in Appendix. \newpage

\section{Worldline supersymmetry\label{secWorldline}}
\subsection{Action with foliation preserving local supersymmetry}
The action for the partonic description of a $p$-brane with a local  worldline supersymmetry we propose is,  with the trace defined in Eq.(\ref{Trace}):
\be
S_{\wSUSY}=\int\!\rd\tau~\Tr\big(\hat{\cL}\big)\,,
\label{actionwSUSY}
\ee
where
\be
\ba{ll}
\hat{\cL}=&\!\!\half D_{\tau}X^{M}D_{\tau}X_{M}-\textstyle{\frac{1}{\,2p!}}
\left\{X^{M_{1}},X^{M_{2}},\cdots,X^{M_{p}}\right\}_{\NB}\left\{X_{M_{1}},X_{M_{2}},\cdots,X_{M_{p}}\right\}_{\NB}\\
{}&+i\half\psi^{M}D_{\tau}\psi_{M}
+i\textstyle{\frac{1}{\,2p!}}\psi^{M_{1}M_{2}\cdots M_{p}}D_{\tau}\psi_{M_{1}M_{2}\cdots M_{p}}\\
{}&-i\textstyle{\frac{1}{(p-1)!}}\psi^{M_{1}M_{2}\cdots M_{p}}\left\{X_{M_{1}},X_{M_{2}},\cdots,X_{M_{p-1}},\psi_{M_{p}}\right\}_{\NB}\\
{}&+i\chi\left(D_{\tau}X^{M}\psi_{M}+\textstyle{\frac{1}{p!}}\left\{X^{M_{1}},X^{M_{2}},\cdots,X^{M_{p}}\right\}_{\NB}
\psi_{M_{1}M_{2}\cdots M_{p}}\right)\,.
\ea
\ee
In addition to the bosonic fields in (\ref{actionbosonic}) which are $X^{M},\varphi,A^{i}_{\tau}$,\,   the above supersymmetric action contains three kinds of fermions: one-form $\psi_{M}$, $p$-form $\psi_{M_{1}M_{2}\cdots M_{p}}$ and one-dimensional gravitino $\chi$.

The action is invariant under the following  foliation preserving diffeomorphism:\footnote{Under the transformations (\ref{fdiff}), (\ref{fsusy}), the corresponding   Lagrangian,   $\cL_{\wSUSY}=\varphi\hat{\cL}$, transforms to a total derivative. } 
\be
\ba{l}
\delta X^{M}=\upsilon^{\lambda}\partial_{\lambda}X^{M}\,,\\
\delta \varphi=\partial_{\lambda}(\upsilon^{\lambda}\varphi)-2\varphi D_{\tau}\upsilon^{\tau}\,,\\
\delta A_{\tau}^{i}=D_{\tau}\upsilon^{i}+
\left(D_{\tau}\upsilon^{\tau}+\upsilon^{\lambda}\partial_{\lambda}\right) A_{\tau}^{~i}\,,\\
\delta\psi^{M}=\upsilon^{\lambda}\partial_{\lambda}\psi^{M}+\half (D_{\tau}\upsilon^{\tau})\psi^{M}\,,\\
\delta\psi^{M_{1}M_{2}\cdots M_{p}}=\upsilon^{\lambda}\partial_{\lambda}\psi^{M_{1}M_{2}\cdots M_{p}}
+\half (D_{\tau}\upsilon^{\tau})\psi^{M_{1}M_{2}\cdots M_{p}}\,,\\
\delta\chi=\upsilon^{\lambda}\partial_{\lambda}\chi+\half(D_{\tau}\upsilon^{\tau})\chi\,,
\ea
\label{fdiff}
\ee
where
\be
\partial_{i}\upsilon^{\tau}=0\,,
\ee
such that in fact, $D_{\tau}\upsilon^{\tau}=\frac{\,\rd \upsilon^{\tau}}{\rd t}$.  Namely while $\upsilon^{i}(\tau,\sigma^{j})$ is an arbitrary  local parameter on the $p$-brane  worldvolume,  
$\upsilon^{\tau}(\tau)$  is arbitrary only over the worldline direction and independent of the spatial  coordinates  $\sigma^{i}$.   From the quantum mechanical point of view, the former generates a gauge symmetry, while  the latter corresponds to the genuine worldline diffeomorphism.   The action is also invariant under  a  foliation preserving local supersymmetry:
\be
\ba{l}
\delta X^{M}=i\psi^{M}\varepsilon\,,\\
\delta\varphi=-2i\varphi\chi\varepsilon\,,\\
\delta A_{\tau}^{i}=0\,,\\
\delta\psi^{M}=D_{\tau}X^{M}\varepsilon\,,\\
\delta\psi^{M_{1}M_{2}\cdots M_{p}}=\left\{X^{M_{1}},X^{M_{2}},\cdots,X^{M_{p}}\right\}_{\NB}\varepsilon
+i\psi^{M_{1}M_{2}\cdots M_{p}}\chi\varepsilon\,,\\
\delta\chi=D_{\tau}\varepsilon+\half(\partial_{i}A^{i}_{\tau}-\varphi^{-1}D_{\tau}\varphi)\varepsilon\,,
\ea
\label{fsusy}
\ee
where $\varepsilon$  is a local fermionic parameter which has arbitrary dependence on the worldline but is  independent of the worldvolume spatial  coordinates,
\be
\partial_{i}\varepsilon=0\,.
\ee
The supersymmetry algebra reads
\be
\delta_{\varepsilon_{1}}\delta_{\varepsilon_{2}}-\delta_{\varepsilon_{2}}\delta_{\varepsilon_{1}}=\delta_{\upsilon}\,,
\ee
where  the right hand side is given by the diffeomorphism parameter,
\be
\ba{cc}
\upsilon^{\tau}=2i\varepsilon_{1}\varepsilon_{2}\,,~~~&~~~\upsilon^{i}=-2i\varepsilon_{1}\varepsilon_{2}A^{i}\,,
\ea
\ee
such that  the  foliation  structure is preserved.   From the quantum mechanical point of view,  the anti-commutator of the one-dimensional local supersymmetry amounts to  a one-dimensional diffeomorphism plus a gauge symmetry, as usual for supersymmetric gauge theories.   Compared to the bosonic action (\ref{actionbosonic}), the action (\ref{actionwSUSY}) lacks the full $(p{+1})$-dimensional worldvolume diffeomorphism, but this is consistent with the no-go theorem  against  the construction of a   worldvolume   supersymmetric action for a $p$-brane of ${p>1}$~\cite{Bergshoeff:1988ui}.\\

\subsection{Action with global worldline supersymmetry}
A  gauge fixed  ($\varphi\equiv 1$ and $\chi\equiv 0$)  action follows:  in terms of the  Filippov-Lie  $p$-bracket,
\be
\ba{ll}
\!\!\cL^{\prime}_{\wSUSY}\!=&\!\!\!\!\half D_{\tau}X^{M}D_{\tau}X_{M}-\textstyle{\frac{1}{\,2p!}}
\left[X^{M_{1}},X^{M_{2}},\cdots,X^{M_{p}}\right]\left[X_{M_{1}},X_{M_{2}},\cdots,X_{M_{p}}\right]\\
{}&\!\!\!\!\!+i\half\psi^{M}D_{\tau}\psi_{M}
+i\textstyle{\frac{1}{\,2p!}}\psi^{M_{1}\cdots M_{p}}D_{\tau}\psi_{M_{1}\cdots M_{p}}
-i\textstyle{\frac{1}{(p-1)!}}\psi^{M_{1}\cdots M_{p{-1}}M_{p}}\left[X_{M_{1}},\cdots,X_{M_{p{-1}}},\psi_{M_{p}}\right].
\ea
\label{globalw}
\ee
The action is clearly invariant under the Filippov-Lie  $\hd$-algebra  gauge transformation~(\ref{Filippovgaugetransformation}),  and further enjoys one 
global  worldline supersymmetry: with a constant parameter $\varepsilon_{0}$,
\be
\ba{l}
\delta X^{M}=i\psi^{M}\varepsilon_{0}\,,\\
\delta A_{\tau}^{i}=0\,,\\
\delta\psi^{M}=D_{\tau}X^{M}\varepsilon_{0}\,,\\
\delta\psi^{M_{1}M_{2}\cdots M_{p}}=[X^{M_{1}},X^{M_{2}},\cdots,X^{M_{p}}]\varepsilon_{0}\,.
\ea
\ee
~\\

Especially,  when $p=1$ \textit{i.e.~}string,  both fermions $\psi^{M}$, $\psi^{M_{1}M_{2}\cdots M_{p}}$ are on the equal footing carrying only one spacetime index,  and there appears an additional  $\SO(2)$ $R$-symmetry in the action.  Consequently the supersymmetry is doubled and the above action reduces, after putting $A^{i}_{\tau}\equiv 0$, to the well-known conformal gauge fixed Ramond-Neveu-Schwarz superstring action~\cite{Green:1987sp}. The case of ${p=2}$ is similar to the BFSS $\cM$-theory matrix model~\cite{Banks:1996vh}, but different points in our action are the types of fermions, the worldline supersymmetry and the full  target spacetime  Lorentz invariance. \\~\newpage

\section{Spacetime supersymmetry\label{secSpacetime}}
\subsection{Action with kappa-symmetry} 
The  spacetime supersymmetric Green-Schwarz covariant  $p$-brane  Lagrangian  reads~\cite{Achucarro:1987nc}
\be
-\sqrt{-\det\!\left(
 \Pi^{M}_{\mu}\Pi_{\nu M}\right)}+\cL_{\WZ}\,,
\label{spacetimeNG}
\ee
where $\Pi^{M}_{\mu}=\partial_{\mu}X^{M}-i\bar{\theta}\Gamma^{M}\partial_{\mu}\theta$ and $\cL_{\WZ}$ corresponds to the Wess-Zumino term necessary for the kappa-symmetry.  The super $p$-brane action exists if and only if the Bose and Fermi degrees of freedom match, such that the  possible values of $p$ and the spacetime dimension $D$ are 
\be
\ba{ll}
p=1~:&~D=3,4,6,10\\
p=2~:&~D=4,5,7,11\\
p=3~:&~D=6,8\\
p=4~:&~D=9\\
p=5~:&~D=10\,.
\ea
\label{pDtable}
\ee
Our partonic reformulation  of the    spacetime supersymmetric  Green-Schwarz $p$-brane Lagrangian    is then:
\be
\cL_{\stSUSY}=\cL(\varphi,A^{i}_{\tau},\Pi^{M}_{\mu})+\cL_{\WZ}\,,
\label{stSUSY}
\ee
where, with $\cV_{ij}:=\Pi^{M}_{i}\Pi_{jM}$,
\be
\cL(\varphi,A^{i}_{\tau},\Pi^{M}_{\mu})=\half\varphi(\Pi^{M}_{\tau}-A^{i}_{\tau}\Pi^{M}_{i})(\Pi_{\tau M}-A^{j}_{\tau}\Pi_{jM})
-\half\varphi^{-1}\det\cV\,.
\ee
In particular,  in a similar fashion to (\ref{potential}), we may write~\cite{Lee:2009ue}
\be
\ba{l}
\varphi^{-2}\det\cV\!=\textstyle{\frac{1}{p!}}\langle \Pi^{M_{1}},\Pi^{M_{2}},\cdots,X^{M_{\hd}}\rangle\langle
\Pi_{M_{1}},\Pi_{M_{2}},\cdots,\Pi_{M_{\hd}}\rangle\,,\\
\langle\Pi^{M_{1}},\Pi^{M_{2}},\cdots,X^{M_{\hd}}\rangle:=\varphi^{-1}\epsilon^{i_{1}i_{2}\cdots i_{{p}}}\Pi_{i_{1}}^{M_{1}}\Pi_{i_{2}}^{M_{2}}\cdots\Pi_{i_{p}}^{M_{p}}\,.
\ea
\ee
The auxiliary fields assume the following  on-shell values,
\be
\ba{lll}
A^{i}_{\tau}&~~\Longrightarrow~~&\hat{A}^{i}_{\tau}:=\Pi^{M}_{\tau}\Pi_{j M}\cV^{-1 ji}\,,\\
\varphi&~~\Longrightarrow~~&\hat{\varphi}:=\sqrt{-\frac{\det\cV}{\,(\Pi^{M}_{\tau}-\hat{A}^{i}_{\tau}\Pi^{M}_{i})(\Pi_{\tau M}-\hat{A}^{j}_{\tau}\Pi_{jM})}\,.}
\ea
\ee
Integrating them  out reduces $\cL(\varphi,A^{i}_{\tau},\Pi^{M}_{\mu})$ to the supersymmetric Nambu-Goto term in (\ref{spacetimeNG}),
\be
\ba{lll}
\cL(\varphi,A^{i}_{\tau},\Pi^{M}_{\mu})&~~\Longrightarrow~~&\cL(\hat{\varphi},\hat{A}^{i}_{\tau},\Pi^{M}_{\mu})=-\sqrt{-\det\!\left(
 \Pi^{M}_{\mu}\Pi_{\nu M}\right)}\,.
\ea
\ee
Furthermore, along with an arbitrary transformation $\delta\Pi_{\mu}^{M}$,  if we let  the auxiliary fields transform as
\be
\ba{l}
\dis{\delta A^{i}_{\tau}=\delta\hat{A}^{i}_{\tau}+\half(\hat{A}^{i}_{\tau}-A^{i}_{\tau})\delta\ln\varphi+(\hat{A}^{k}_{\tau}-A^{k}_{\tau})\Pi^{M}_{k}\delta\Pi_{j M}\cV^{-1 ji}\,,}\\
\dis{\delta\varphi=\frac{2\varphi^{2}}{(\hat{\varphi}+\varphi)\hat{\varphi}}\delta\hat{\varphi}+
\frac{(\hat{\varphi}-\varphi)\varphi}{\hat{\varphi}+\varphi}\delta\ln\det\cV\,,}
\ea
\ee
the variation of $\cL(\varphi,A^{i}_{\tau},\Pi^{M}_{\mu})$ becomes independent of the auxiliary fields and, moreover, coincides with  that of the spacetime supersymmetric  Nambu-Goto term:
\be
\delta\cL(\varphi,A^{i}_{\tau},\Pi^{M}_{\mu})=\delta\cL(\hat{\varphi},\hat{A}^{i}_{\tau},\Pi^{M}_{\mu})=-\delta\sqrt{-\det\!\left(
 \Pi^{M}_{\mu}\Pi_{\nu M}\right)}\,.
\ee
Therefore, all the  symmetries  of the  Green-Schwarz super  $p$-brane Lagrangian (\ref{spacetimeNG}) survive in our  partonic reformulation (\ref{stSUSY}), which include the spacetime supersymmetry, the spacetime Lorenz symmetry, the  kappa-symmetry and the worldvolume diffeomorphism.\footnote{See  \cite{Lee:2009ue,Kamani:2009wg} for similar works,  and 
recall the general phenomenon  that no symmetry is lost under an  arbitrary  reformulation of a given  action by   auxiliary fields~\cite{Bergshoeff:1988ui}.}\\

\subsection{Action with global  spacetime supersymmetry}
The light-cone gauge fixed actions are   ready to be read-off from an earlier work
by Bergshoeff, Sezgin, Tanii and Townsend~\cite{Bergshoeff:1988hw}.
In its appendix the authors listed  light-cone gauge fixed  supersymmetric actions for  various $p$-branes in diverse spacetime dimensions. Utilizing the identity (\ref{potential}), in terms of Filippov-Lie algebra  $p$-bracket,  their light-cone gauge fixed spacetime supersymmetric $p$-brane  actions can be rewritten   in a compact form:
\be
\cL^{\prime}_{\stSUSY}=\half (D_{t}X^{I})^{2}-\textstyle{\frac{1}{2p!}}
\left[X^{I_{1}},X^{I_{2}},\cdot\cdot,X^{I_{p}}\right]^{2}+i\half\bar{\theta}D_{t}\theta+
\textstyle{\frac{1}{2(p-1)!}}\bar{\theta}\Gamma^{I_{1}I_{2}\cdots I_{p-1}}\left[X_{I_{1}},\cdots,X_{I_{p-1}},\theta\right]\,.
\label{globalst}
\ee
Here the spacetime index $I$ runs from one to $D-2$ with  the  possible values of $p$ and $D$ in (\ref{pDtable}). For details of the supersymmetry transformation we refer to Ref.\cite{Bergshoeff:1988hw}.\footnote{See also Ref.\cite{Furuuchi:2009ax} for a zero-dimensional analogy.}\newpage

\section{Discussion\label{secDiscussion}}
 In this paper we have constructed   supersymmetric extensions of   a bosonic $p$-brane action which reformulates the Nambu-Goto action as an  interacting  multi-particle action with    Filippov-Lie $p$-algebra gauge  symmetry.   We  obtained   a worldline  supersymmetric  action by postulating, among others,     a $p$-form fermion.   We also derived  a    spacetime supersymmetric  Green-Schwarz  extension    from   the preexisting    kappa-symmetric action. 

Compared to the ordinary Lie algebra, one limited feature of Filippov-Lie $p$-algebra for $p\geq 3$ is that,     finite dimensional irreducible Filippov-Lie  algebra is essentially unique, \textit{i.e.} $\so(p{+1})$~\cite{FigueroaO'Farrill:2002xg,Papadopoulos:2008sk,Gauntlett:2008uf,Papadopoulos:2008gh}. Consequently  there is no arbitrary tunable parameter as for the number of finite degrees of freedom. One should deal with infinite dimensional Filippov-Lie $p$-algebras or $\so(p{+1})$.  In the latter  case  the  $p$-brane   corresponds to a fuzzy $p$-sphere.

For ${p=2}$, the situation is different.  We may safely adopt matrices of arbitrary  size. With the usual matrix commutator, our  actions read from (\ref{globalw}),
\be
\cL_{\wSUSY}=\half D_{\tau}X^{M}D_{\tau}X_{M}-\textstyle{\frac{1}{4}}
\left[X^{M},X^{N}\right]^{2}+i\half\psi^{M}D_{\tau}\psi_{M}
+i\textstyle{\frac{1}{4}}\psi^{MN}D_{\tau}\psi_{MN}
-i\psi^{MN}\left[X_{M},\psi_{N}\right]\,,
\label{wM2}
\ee
and from (\ref{globalst}),
\be
\cL_{\stSUSY}=\half D_{t}X^{I}D_{t}X_{I}-\textstyle{\frac{1}{4}}\left[X^{M},X^{N}\right]^{2}
+i\half\bar{\theta}D_{t}\theta+
\textstyle{\frac{1}{2}}\bar{\theta}\Gamma^{I}\left[X_{I},\theta\right]\,.
\ee
Of course, the latter corresponds to the well-known BFSS $\cM$-theory matrix model~\cite{Banks:1996vh} where the fermion is a  target spacetime spinor.  In analogy with the equivalence between the   RNS and the GS superstring actions, it is crucial to check the connection between the above two actions for M2-brane.

In the case of ${p=0}$, with vanishing  Nambu bracket,  our worldline supersymmetric  action (\ref{actionwSUSY})   reduces to the well-known action for a massless supersymmetric particle (see \textit{e.g.~}\cite{Gomis:1994he}). Our action   then corresponds to  non-Abelian or  Filippov-Lie algebra generalization of it, where a `mass' term appears as the square of Nambu bracket.   We recall the known difficulty  that a massive point-particle does not allow a worldline supersymmetric extension,  as there is no supersymmetric counter part to its mass term,
\be
\dis{S=\half\int{\rm d}\tau \left( e \partial_{\tau}X^{M}\partial_{\tau}X_{M}-e^{-1}m^{2}\right)\,.}
\ee
While a compact $p$-brane should look like a point-particle at far distance,   our result  delivers  a novel way of introducing a supersymmetric mass term: the  Filippov-Lie algebra based interaction of the  partons of the compact $p$-brane gives rise to the mass term. 
Namely,  mass originates from the internal  interaction, like a proton made of light  quarks.  Further, we expect that  the constraint form the  equation of motion for $\chi$,\footnote{\textit{c.f.~}\cite{Bonelli:2005ti,Bonelli:2008kh}  
and some BPS equations in the Bagger-Lambert-Gustavsson theory~\cite{Jeon:2008bx}.}
\be  
\psi^{M}D_{\tau}X_{M}+\textstyle{\frac{1}{p!}}\psi^{M_{1}M_{2}\cdots M_{p}}\left\{X_{M_{1}},X_{M_{2}},\cdots,X_{M_{p}}\right\}_{\NB}
=0\,,
\ee
also leads, after quantization, to a certain massive  Dirac equation.   We leave the quantization of the supersymmetric $p$-brane in our formulation  for future work. 
In the present paper we focused on the partonic description of a super $p$-brane.  Since the aforementioned   no-go theorem  prohibits   the construction of a worldvolume  supersymmetric $p$-brane action for  $p{>1}$,  the other  alternative possibility worth trying   is  the supersymmetric extensions  of the `multi-string description of a  $p$-brane,'   starting from the  bosonic  action~\cite{Park:2008qe},
\be
\ba{l}
\dis{S_{\multistring}=\int\rd^{2}\tau~\Tr\left(\sqrt{-h\,}\cL_{\multistring}\right)\,,}
~~~~~~~\dis{\Tr:=\int\rd^{p{-1}}\sigma\,,}\\
\cL_{\multistring}=-e^{-\phi}h^{ab}D_{a}X^{M}D_{b}X_{M}
-\textstyle{\frac{1}{4}}e^{\phi}\det V+e^{-\phi}\,,
\ea
\ee 
where $a,b=0,1$ are the two-dimensional  worldsheet indices~\cite{Kanghoon}. The resulting action may correspond to a RNS version of the well-known matrix string action~\cite{Dijkgraaf:1997vv}.

~\\
~\\
\noindent\textbf{Acknowledgements}\\
We  thank  Seungjoon Hyun and Corneliu Sochichiu  for helpful comments. The work is supported by the National Research Foundation of Korea(NRF) grant funded by the Korea government(MEST)
through the Center for Quantum Spacetime(CQUeST) of Sogang University with grant number 2005-0049409, and   by the National Research Foundation of Korea(NRF) grant funded by the Korea government(MEST) (No. 2009-0083765).\newpage

\appendix

\section{Useful relations\label{secAppendix}}
Here we write some useful identities:
\be
\ba{l}
\{X^{[M_{1}},\cdots,X^{M_{p}}\}_{\NB}\,\partial_{i}X^{M_{p+1}]}=0~~~~~~\mbox{for~\,arbitrary~~~}i\,,\\
\{X_{[M_{1}},\cdots,X_{M_{p-1}},\upsilon^{i}\}_{\NB}\partial_{i}X_{M_{p}]}
=\textstyle{\frac{1}{p}}\partial_{i}\upsilon^{i}\{X_{M_{1}},\cdots,X_{M_{p}}\}_{\NB}\,,
\ea
\label{useful}
\ee

\be
\ba{l}
\varphi^{-2}\det V V^{-1ij}\partial_{i}Y\partial_{j}Z=\textstyle{\frac{1}{(\hd-1)!}}
\{X^{M_{1}},X^{M_{2}},\cdots,X^{M_{\hd-1}},Y\}_{\NB}
\{X_{M_{1}},X_{M_{2}},\cdots,X_{M_{\hd-1}},Z\}_{\NB}\,,\\
\varphi^{-1}\partial_{i}\!\left(\varphi^{-1}\det V V^{-1ij}\partial_{j}Y\right)=
\textstyle{\frac{1}{(\hd-1)!}}\{X^{M_{1}},X^{M_{2}},\cdots,X^{M_{\hd-1}},
\{X_{M_{1}},X_{M_{2}},\cdots,X_{M_{\hd-1}},Y\}_{\NB}\}_{\NB}\,,
\ea
\label{useful2}
\ee

\be
\ba{l}
\det V=\textstyle{\frac{1}{p!}}\epsilon^{r_{1}r_{2}\cdots r_{p}}
\epsilon^{s_{1}s_{2}\cdots s_{p}}
V_{r_{1}s_{1}}V_{r_{2}s_{2}}\cdots V_{r_{p}s_{p}}\,,\\
\frac{\partial^{n}\det V}{\partial V_{i_{1}j_{1}}\partial V_{i_{2}j_{2}}
\cdots\partial V_{i_{n}j_{n}}}=
\textstyle{\frac{1}{(p-n)!}}\epsilon^{i_{1}\cdots i_{n} r_{1}\cdots r_{p-n}}\epsilon^{j_{1}\cdots j_{n} s_{1}\cdots s_{p-n}}
V_{r_{1}s_{1}}\cdots V_{r_{p-n}s_{p-n}}\,.
\ea
\label{useful3}
\ee
In particular,
\be
\ba{l}
 V^{-1ij}\det V=\textstyle{\frac{1}{(p-1)!}}\epsilon^{ir_{1}\cdots r_{p-1}}\epsilon^{j s_{1}\cdots s_{p-1}}
V_{r_{1}s_{1}}\cdots V_{r_{p-1}s_{p-1}}\,,\\
\left(V^{-1ij}V^{-1kl}-V^{-1ik}V^{-1lj}\right)\det V=\textstyle{\frac{1}{(p-2)!}}\epsilon^{ikr_{1}\cdots r_{p-2}}\epsilon^{jl s_{1}\cdots s_{p-2}}
V_{r_{1}s_{1}}\cdots V_{r_{p-2}s_{p-2}}\,.
\ea
\ee
~\\

Under the worldvolume diffeomorphism (\ref{diffgend1}), 
\be
\ba{rl}
{\delta_{\diff} D_{\tau}X^{M}=}&
\textstyle{\frac{1}{(\hd-1)!}}
\{X^{N_{1}},\cdots,X^{N_{\hd-1}},\upsilon^{\tau}\}_{\NB}
\{X_{N_{1}},\cdots,X_{N_{\hd-1}},X^{M}\}_{\NB}\\
{}&+\left(D_{\tau}\upsilon^{\tau}+\upsilon^{\mu}\partial_{\mu}\right) D_{\tau}X^{M}\,,\\
\delta_{\diff}\left\{X^{M_{1}},\cdots,X^{M_{\hd}}\right\}_{\NB}=&\displaystyle{\sum_{k=1}^{p}\,D_{\tau}X^{M_{k}}
\left\{X^{M_{1}},\cdots,X^{M_{k-1}},\upsilon^{\tau},X^{M_{k+1}},\cdots,X^{M_{\hd}}\right\}_{\NB}}\\
{}&+\left(D_{\tau}\upsilon^{\tau}+\upsilon^{\mu}\partial_{\mu}\right)\!\left\{X^{M_{1}},\cdots,X^{M_{\hd}}\right\}_{\NB}\,.
\ea
\ee

\newpage


\begin{thebibliography}{99}


\bibitem{Achucarro:1987nc}
  A.~Achucarro, J.~M.~Evans, P.~K.~Townsend and D.~L.~Wiltshire,
  Phys.\ Lett.\  B {\bf 198} (1987) 441.

\bibitem{Deser:1976rb}
  S.~Deser and B.~Zumino,
  Phys.\ Lett.\  B {\bf 65} (1976) 369.


\bibitem{Brink:1976sc}
  L.~Brink, P.~Di Vecchia and P.~S.~Howe,
  Phys.\ Lett.\  B {\bf 65} (1976) 471.





\bibitem{Howe:1977hp}
  P.~S.~Howe and R.~W.~Tucker,
  J.\ Phys.\ A  {\bf 10} (1977) L155.
  
  
\bibitem{Bergshoeff:1988ui}
  E.~Bergshoeff, E.~Sezgin and P.~K.~Townsend,
  Phys.\ Lett.\  B {\bf 209} (1988) 451.

\bibitem{Park:2008qe}
  J.-H.~Park and C.~Sochichiu,
  Eur. Phys. J. C {\bf 64} (2009) 161-166
  [arXiv:0806.0335  hep-th].
  
  
\bibitem{Nambu:1973qe}
  Y.~Nambu,
  Phys.\ Rev.\ D {\bf 7} 2405 (1973).
  
   
  
\bibitem{Henneaux:1983um}
  M.~Henneaux,
  Phys.\ Lett.\  B {\bf 120} (1983) 179.

  
\bibitem{Schild:1976vq}
  A.~Schild,
 Phys.\ Rev.\  D {\bf 16} (1977) 1722.


\bibitem{Hoppe}
J.~Hoppe,
MIT PhD thesis 1982 and Elem. Part. Res. J. (Kyoto) {\bf 80} (1989) 145;\\
  J.~Hoppe,
  Helv.\ Phys.\ Acta {\bf 70} (1997) 302
  [arXiv:hep-th/9602020].



\bibitem{Bergshoeff:1988hw}
  E.~Bergshoeff, E.~Sezgin, Y.~Tanii and P.~K.~Townsend,
  Annals Phys.\  {\bf 199} (1990) 340.




\bibitem{BL}
  J.~Bagger and N.~Lambert,
  Phys.\ Rev.\  D {\bf 75} (2007) 045020
  [arXiv:hep-th/0611108];\\
  J.~Bagger and N.~Lambert,
  Phys.\ Rev.\  D {\bf 77} (2008) 065008
  [arXiv:0711.0955 [hep-th]];\\
  J.~Bagger and N.~Lambert,
  JHEP {\bf 0802} (2008) 105
  [arXiv:0712.3738 [hep-th]].

\bibitem{Gustavsson:2007vu}
  A.~Gustavsson,
  Nucl.\ Phys.\  B {\bf 811}, 66 (2009)
  [arXiv:0709.1260 [hep-th]].



\bibitem{Green:1987sp}
  M.~B.~Green, J.~H.~Schwarz and E.~Witten,
  ``SUPERSTRING THEORY. VOL. 1: INTRODUCTION,''
{\it  Cambridge, Uk: Univ. Pr.} ( 1987).





\bibitem{Banks:1996vh}
  T.~Banks, W.~Fischler, S.~H.~Shenker and L.~Susskind,
  Phys.\ Rev.\  D {\bf 55} (1997) 5112
  [arXiv:hep-th/9610043].
  
\bibitem{Myers:1999ps}
  R.~C.~Myers,
  JHEP {\bf 9912} (1999) 022
  [arXiv:hep-th/9910053].

  
\bibitem{Yang:1998qd}
  H.~S.~Yang, I.~Kim and B.~H.~Lee,
  Phys.\ Rev.\  D {\bf 58} (1998) 085018
  [arXiv:hep-th/9806112].



\bibitem{Lindstrom:1988az}
  U.~Lindstrom and M.~Rocek,
  Phys.\ Lett.\  B {\bf 218} (1989) 207.


\bibitem{Castro:2002ra}
  C.~Castro,
  Phys.\ Lett.\  B {\bf 559} (2003) 74
  [arXiv:hep-th/0212022].


\bibitem{NambuGoto}
Y.~Nambu, Lectures at the Copenhagen symposium, 1970;
  T.~Goto,
  Prog.\ Theor.\ Phys.\  {\bf 46} (1971) 1560.





\bibitem{Takhtajan:1993vr}
  L.~Takhtajan,
  Commun.\ Math.\ Phys.\  {\bf 160} (1994) 295
  [arXiv:hep-th/9301111];\\
  H.~Awata, M.~Li, D.~Minic and T.~Yoneya,
  JHEP {\bf 0102} (2001) 013
  [arXiv:hep-th/9906248].
  
  
  


\bibitem{n-Lie}
V. T. Filippov, ``n-Lie algebras,"  Sib. Mat. Zh., 26, No 6, 126-140 (1985).


\bibitem{FigueroaO'Farrill:2002xg}
  J.~M.~Figueroa-O'Farrill and G.~Papadopoulos,
  arXiv:math/0211170.


\bibitem{Papadopoulos:2008sk}
  G.~Papadopoulos,
  JHEP {\bf 0805} (2008) 054
  [arXiv:0804.2662 [hep-th]].

\bibitem{Gauntlett:2008uf}
  J.~P.~Gauntlett and J.~B.~Gutowski,
  arXiv:0804.3078 [hep-th].


\bibitem{Papadopoulos:2008gh}
  G.~Papadopoulos,
  Class.\ Quant.\ Grav.\  {\bf 25} (2008) 142002
  [arXiv:0804.3567 [hep-th]].
  
  

\bibitem{Lee:2009ue}
  K.~Lee and J.-H.~Park,
  JHEP {\bf 0904} (2009) 012
  [arXiv:0902.2417 [hep-th]].

\bibitem{Kamani:2009wg}
  D.~Kamani,
  arXiv:0904.2721v3 [hep-th].


\bibitem{Furuuchi:2009ax}
  K.~Furuuchi and D.~Tomino,
  JHEP {\bf 0905} (2009) 070
  [arXiv:0902.2041 [hep-th]].


  
\bibitem{Gomis:1994he}
  J.~Gomis, J.~Paris and S.~Samuel,
  Phys.\ Rept.\  {\bf 259} (1995) 1
  [arXiv:hep-th/9412228].





\bibitem{Bonelli:2005ti}
  G.~Bonelli and M.~Zabzine,
  JHEP {\bf 0509} (2005) 015
  [arXiv:hep-th/0507051].




\bibitem{Bonelli:2008kh}
  G.~Bonelli, A.~Tanzini and M.~Zabzine,
  Phys.\ Lett.\  B {\bf 672} (2009) 390
  [arXiv:0807.5113 [hep-th]].




\bibitem{Jeon:2008bx}
  I.~Jeon, J.~Kim, N.~Kim, S.~W.~Kim and J.-H.~Park,
  JHEP {\bf 0807} (2008) 056
  [arXiv:0805.3236 [hep-th]].



\bibitem{Kanghoon}
K. Lee and J.-H. Park \textit{in preparation.}
  
\bibitem{Dijkgraaf:1997vv}
  R.~Dijkgraaf, E.~P.~Verlinde and H.~L.~Verlinde,
  Nucl.\ Phys.\  B {\bf 500} (1997) 43
  [arXiv:hep-th/9703030].

  

\end{thebibliography}
\end{document}